\documentclass[12pt,a4paper]{revtex4}
\usepackage{color}
\usepackage{graphicx}
\usepackage{subfig}
\usepackage{bm}

\begin{document}
\title{\bf Turbulence model for simulation of the flame\\
  front propagation in SNIa}

\author{Glazyrin S.I.}
\email{glazyrin@itep.ru}
\affiliation{
  Institute for Theoretical and Experimental Physics, Moscow, Russia
}
\altaffiliation{
  Present address: All-Russia Research Institute of Automatics, Moscow, Russia
}

\begin{abstract}
  Turbulence significantly influences the dynamics of flame in
  SNIa. The large Reynolds number makes impossible the direct numerical
  simulations of turbulence, and different models of
  turbulence have to be used. Here we present the simulations with the
  $k$-$\epsilon$ model. The turbulence is generated by the RTL instability
  and crucially influences flame front velocity, resulting in
  $v_{\rm flame}\sim 300$ km/s. The model reproduces turbulent
  properties in low-dimensional simulations and can be used for the
  low-cost studies.
\end{abstract}

\maketitle

\section{Introduction}

The problem of a nuclear flame propagation in the SNIa is still
controversial and is one of the fundamental questions in astrophysics
and the theory of burning. These flashes are considered to be the thermonuclear
explosions of a white dwarf close to Chandrasekhar limit in binary systems.
There exist three popular scenarios of SNIa events: single--degenerate,
double--degenerate and sub--Chandrasekhar explosions (see,
e.g. \cite{HillebrandtNiemeyer_astro_ph_0006305}).
The first depends crucially on flame physics: to meet observations it
requires two stages of burning propagation, slow burning (the deflagration)
and the detonation. The effective mechanism of deflagration to
detonation transition (DDT) should exist \cite{KhokhlovEtAl_ApJ_1997, LisewskiEtAl_ApJ_2000, SeitenzahlEtAl_MNRAS_2013} and is an essential part of the
single--degenerate scenario considered here, when the progenitor of the
explosion is a white dwarf (WD) in a binary system with a
non-degenerate companion star.

The flame in conditions of a white dwarf is negligibly thin compared
to any other spatial scale \cite{TimmesWoosley_ApJ_1992}. Such a flame is subject to many
instabilities
\cite{HillebrandtNiemeyer_astro_ph_0006305,NiemeyerWoosley_ApJ_1997,Glazyrin_AstronLett_2013}. The
effects of such instabilities are in changing the character of the flame
propagation (like the change of its velocity) or turbulization of
medium (and influence on flame through turbulence). The latter is considered in this
paper. The flame velocity acceleration close to the speed of sound is an
essential part of DDT and turbulence could do it effectively.

Though several effects lead to turbulization, only one instability is
considered in this paper as it plays the dominating role in SNIa. This
is the instability of the surface separating two mediums with
different densities in gravitational field. For the non-interacting
mediums it is called the Rayleigh--Taylor instability, and for flames
it was first considered by Landau in \cite{Landau_1944} (the famous
Landau--Darries instability was also introduced in that paper), so we
will use below the term Rayleigh--Taylor--Landau instability (RTL) for flames.
The surface can be unstable when density gradient and gravitational acceleration
satisfy: ${\bf g}\nabla\rho<0$, what is true for a flame spreading
outwards the centre of a star.

The turbulent flames in SNIa were considered in a number of papers.
In the paper \cite{NiemeyerHillebrandt_ApJ_1995} the
subgrid-scale model for turbulence was implemented, the main source of
turbulence was the RTL instability, as in our work. It was shown that
the turbulence increases flame velocity to $\sim 2$\% of the sound
speed. This model of turbulence was significantly updated in
papers \cite{SchmidtEtAl_AstronAstrophys_2006a,
  SchmidtEtAl_AstronAstrophys_2006b}. With this new model a full 3D
simulation of a star is presented in \cite{Ropke_ApJ_2007}. It was
shown that deflagration--to--detonation transition could be fully
flame driven, as there are non-negligible probabilities of turbulent
velocities $v'>10^3$ km/s (for details see \cite{Ropke_ApJ_2007}).
Another approach is presented in
\cite{WoosleyKersteinEtAl_ApJ_2009}, where the authors used
the sophisticated 1D semi--empirical model for turbulent flames called
Linear--Eddy--Model (LEM). The authors succeeded in DDT explanation
in conditions on border of two regimes of turbulent flame propagation.
Also we should mention works like \cite{AspdenEtAl_ApJ_2008}, where
the microphysics of turbulence--flame interaction is considered.

We use here the semi--empirical $k$-$\epsilon$ model to calculate
turbulent burning and turbulent parameters. The model accurately
simulates Rayleigh--Taylor (RT) and Kelvin--Helmholtz (KH) mixing
processes. The benefit of such a model is that it can correctly
reproduce 3D properties of turbulence in low-dimensional
simulations. In this paper the 1D model of turbulence generation by
the RTL instability at the flame interface together with the impact of
turbulence on the flame is considered. A similar to our model was
proposed in \cite{SimonenkoEtAl_AstronLett_2007} for the problems of
thermonuclear burning on a surface of neutron stars.
The Section \ref{sec:model} presents the model of turbulence, the
setup of the problem for WD together with numerical approach used is
presented in Section \ref{sec:problem}, Section \ref{sec:results}
contains results of simulations, the discussion is presented in
Section \ref{sec:conclusions}. 

\section{The model of turbulent flame}
\label{sec:model}

The class of turbulence models we are using was initially proposed in
paper \cite{JonesLaunder_IntJHeatMassTranf_1972}. These models meet
several requirements: they should
reproduce Navier--Stokes equations when quantities that
characterise turbulence are set to zero; and also reproduce
the leading terms in Navier--Stokes equations for the Reynolds number ${\rm
  Re}\rightarrow\infty$. The simplest procedure to
satisfy these requirements is to build a model by averaging exact
hydrodynamic equations, and try to close the obtained system at some
level of correlators.
The considered model is a Reynolds-averaged model is contrast to
large eddy simulation model by
\cite{SchmidtEtAl_AstronAstrophys_2006a}, which is based on filtering.

The system of hydrodynamic equations with burning looks like:
\begin{equation}
  \label{eq:hydro_rho}
  \partial_t\rho + \partial_i(\rho v_i)=0,
\end{equation}
\begin{equation}
  \label{eq:hydro_rhov}
  \partial_t(\rho v_i)+\partial_j(\rho v_i v_j)+\partial_ip=\partial_j\tau_{ij},
\end{equation}
\begin{equation}
  \label{eq:hydro_E}
  \partial_t(\rho e)+\partial_i\left(\rho e v_i\right)+p\partial_i v_i+\partial_iQ_i=
  \tau_{ij}\partial_jv_i+\dot{S},
\end{equation}
here $\rho$ -- is the density of the medium, $v_i$ -- velocity, $p$ --
pressure, $e$ -- internal energy per unit mass, $\tau_{ij}$ is the
viscous tensor, $Q_i$ -- heat flow, $\dot{S}$ energy generation by
flame.

We will not provide explicit expressions for $Q_i$ and
$\dot{S}$ here as they are not required in the paper. 
But the dynamics of the deflagration flame for laminar flows is
determined by these two processes:
thermoconductivity and energy generation. If $\kappa$ is the
coefficient of thermoconductivity (defined as $Q_i=-\kappa\partial_i
T$), $q$ -- caloricity, then the flame
thickness $\delta$ satisfies
\begin{equation}
  \label{eq:tau_flame}
  \frac{\delta^2}{\kappa}=\frac{q}{\dot{S}}=\tau_{\rm flame}.
\end{equation}
This equality means that for the flame to exist the energy should be
transferred through its thickness to heat the next layer on a timescale of
energy generation \cite{ll6}.

We consider the case when the burning timescale is much smaller than
the turbulent and hydrodynamic
one's (further we will provide an exact criterion). It means that we can
introduce the concept of flame as a thin surface that separates burned
and unburned matter. In this case we can exclude the
thermoconductivity from
Eqns. (\ref{eq:hydro_rho})--(\ref{eq:hydro_E}). But the flame will now
be moved ``by hands'': we should set its normal velocity $u_n$ (as a
function of matter state) and energy generation on the front. In general 3D case
the equation that describes its dynamics reads, e.g., as \cite{ReineckeHillebrandtEtAl_AstronAstrophys_1999}:
\begin{equation}
  \label{eq:G_eq}
  \partial_t G+(v\nabla) G = u_n({\bf x},t)|\nabla G|,
\end{equation}
where $G({\bf x}, t)$ is a level-function:
\begin{eqnarray}
  G<0:&\quad\mbox{unburned matter},\\
  G>0:&\quad\mbox{burned matter}.\nonumber
\end{eqnarray}
For such flame model its normal flame velocity should be pre-calculated in the full--physics
hydrodynamic simulations like \cite{TimmesWoosley_ApJ_1992,
 GlazyrinBlinnikovDolgov_2013}. In the scope of this paper we use the approximation formulas
from \cite{TimmesWoosley_ApJ_1992}. 

The turbulence is characterized by the existence of a cascade: a
interval in the space of wave-numbers with a universal scaling law
where energy is transferred from the large
scale to the dissipation scale. The turbulent pulsations $v'$ depend on the
spatial scale $l$. We can define the Gibson scale $l_G$ as:
\begin{equation}
  \label{eq:l_G_def}
  v'(l_G)=u_n.
\end{equation}
The physical meaning of the scale is that a turbulence does not
influence on a flame on spatial scales $l<l_G$ and affects it on scales $l>l_G$.
The regime of turbulent burning we are considering (the flamelet regime) is described in
terms of the Karlovitz number:
\begin{equation}
  \label{eq:Ka_def}
    {\rm Ka} \equiv \left(\frac{\delta}{l_G}\right)^{1/2}\ll 1.
\end{equation}

To build a model of a turbulence we average hydrodynamic equations.
The following rules are used. The Reynolds averaging
\begin{equation}
  \overline{A}({\bf x},t)=\frac{1}{T}\int\limits_{-T/2}^{T/2}A({\bf x},t+\tau)d\tau,
\end{equation}
where $T$ is the characteristic timescale of turbulent pulsations. For
compressible fluids it is better to use the Favre averaging,
\begin{equation}
  \label{eq:favre_averaging}
  \tilde{A}=\frac{\overline{\rho A}}{\overline{\rho}},
\end{equation}
for some quantities, namely: $v$, $e$. The other quantities, $\rho$, $p$, $\tau_{ij}$,
will be averaged by Reynolds rule. Every quantity can be split into averaged
and pulsational parts:
\begin{equation}
  A=\overline{A}+A'=\tilde{A}+A'',\quad A'=A''+\frac{\overline{\rho' A'}}{\overline{\rho}}.
  \label{eq:Favre_Reyn}
\end{equation}
The latter equality can be deduced from definitions.

After working on hydrodynamic equations we obtain (for details see
\cite{YanilkinStatsenkoKozlov, BHRZ}):
\begin{equation}
  \label{eq:hydro_rho_av}
  \partial_t\overline{\rho} + \partial_i(\overline{\rho} \tilde{v}_i)=0,
\end{equation}
\begin{equation}
  \label{eq:hydro_rhov_av}
  \partial_t(\overline{\rho}\tilde{v}_i)+\partial_j(\overline{\rho}\tilde{v}_i \tilde{v}_j)+
  \partial_j R_{ij}+\partial_i\overline{p}=\partial_j\overline{\tau}_{ij},
\end{equation}
\begin{equation}
  \label{eq:hydro_E_av}
  \partial_t(\overline{\rho}\tilde{e})+\partial_i\left(\overline{\rho} \tilde{e} \tilde{v}_i\right)+
  \partial_i\left(\overline{\rho e''v_i''}\right)+
  \overline{p}\partial_i\overline{v}_i+\overline{p'\partial_i v_i'}=
  \overline{\tau}_{ij}\partial_j\tilde{v}_i+\overline{\tau_{ij}\partial_jv''_i}
  +\overline{\dot{S}},
\end{equation}
with definitions $R_{ij}\equiv\overline{\rho v''_iv''_j}$,
$R_{ij..k}\equiv\overline{\rho v''_iv''_j...v''_k}$ ($R_{ij}$ is a
Reynolds tensor, it describes the momentum transfer by turbulence).

The goal of a turbulence model is to calculate the unknown terms,
second and higher order correlators of pulsational quantities,
$R_{ij}$, $\overline{\rho e'' v_i''}$, etc.

The general idea of $k$-$\epsilon$ models is to introduce two
additional dynamical quantities: the energy of turbulent
pulsations (here $v''$ characterises pulsations on the scale of
a turbulent energy generation, compare Eq.~(\ref{eq:l_G_def}) and the text above it)
and its dissipation:
\begin{equation}
  k\equiv\frac{1}{\overline{\rho}}\frac{\overline{\rho (v''_i)^2}}{2},\quad
  \epsilon\equiv \frac{1}{\overline{\rho}}\overline{\tau_{ij}'\partial_j v_i'}.
\end{equation}
These quantities define the turbulent timescale
$\tau_T=k/\epsilon$. It introduces the turbulent diffusion coefficient
\begin{equation}
  D\sim v''^2\tau_T\sim \frac{k^2}{\epsilon}.
\end{equation}
The diffusion coefficient makes it possible to calculate turbulent
averages $\overline{v''A'}$ with the ``gradient approximation'' \cite{BelenkiiFradkin_1965}:
\begin{equation}
  \label{eq:grad_approx}
  \overline{v_i''A'}\sim -D\partial_i A.
\end{equation}
When applying this approximation, different constants of
proportionality are used for different physical quantities $A$.

The Reynolds tensor with this approximation (with symmetry properties)
is
\begin{equation}
  \label{eq:Rij}
    R_{ij}=-\rho D \left(\partial_i v_j+\partial_j v_i-\frac{2}{3}\delta_{ij}\partial_lv_l\right)
    +\frac{2}{3}\rho k\delta_{ij},
\end{equation}
the first term is the turbulent viscosity, and the second one is the turbulent pressure.

Let us deduce the equation for the turbulent energy $k$. From
(\ref{eq:hydro_rhov}) and (\ref{eq:hydro_rhov_av}) we obtain exact
equation for $k$:
\begin{equation}
  \partial_t(\overline{\rho}k)+\partial_j(\overline{\rho}\tilde{v}_j k)+\frac{1}{2}\partial_j R_{iij}+
  R_{ij}\partial_j\tilde{v}_i=-\overline{v_i''\partial_ip}+\overline{v_i''\partial_j\tau_{ij}}.
  \label{eq:k_exact}
\end{equation}
Let us concisely consider different terms in this equation:
$-R_{ij}\partial_j\tilde{v}_i$ is a shear turbulence generation term $G_1$.
Another generation term appears from
\begin{equation}
  -\overline{v_i''\partial_i p}=-\overline{v_i''}\partial_i\overline{p}-
  \overline{v_i''\partial_i p'}=\frac{\overline{\rho' v_i'}}{\overline{\rho}}\partial_i p-
  \overline{v_i''\partial_i p'},
\end{equation}
here the second term in the RHS is omitted (for low-Mach flows as is
$\sim {\rm Ma}^3$) and the generation term is
\begin{equation}
  G_2\equiv\frac{\overline{\rho' v_i'}}{\overline{\rho}}\partial_i p.
\end{equation}
The expression $-\partial_j R_{iij}/2$ is approximated with the gradient rule as a diffusion term
$\partial_j (c_k\rho D\partial_j
k)$. The term $\overline{v_i''\partial_j\tau_{ij}}$ is $\rho\epsilon$ for
high-Reynolds-number flows. So finally the equation for turbulent
energy is (in final model equations we will drop the notation of averaging):
\begin{equation}
  \partial_t(\rho k)+\partial_i(\rho k v_i)=
  G_1+G_2-\rho\epsilon+\partial_i(\rho c_k D\partial_i k).
\end{equation}

Analogously in Eq.~(\ref{eq:hydro_E_av})
$\overline{\tau_{ij}\partial_j v_i''}$ is approximated as $\rho\epsilon$, terms
$\overline{p'\partial_i v_i'}$ and
$\overline{\tau}_{ij}\partial_j\tilde{v}_i$ are omitted (the first
term for low-Mach flows and from the energy conservation law (see
further); the second is the collisional viscosity, it
is negligible in comparison with the turbulent viscosity),
$\overline{\rho e''v_i''}$ is turbulent thermoconductivity
$Q_i^T=-c_e\rho D\partial_i e$. In the term
$\overline{p}\partial_i\overline{v}_i$ the velocity should be replaced
by Favre average with Eq.~(\ref{eq:Favre_Reyn}), it leads to additional
term $\partial_i(pa_i)$ in the final equation ($a_i\equiv
\overline{\rho'v'}/\overline{\rho}$). We should note that the
final variant of the equation for internal energy $e$ should be consistent with the equation
for $\rho v^2/2$ and $\rho k$ to conserve energy, what is true with proposed earlier
approximations.

The exact equation for $\epsilon$ contains a lot of complex terms that
are not easily approximated in the considered framework, so usually
it is written similar to $k$-equation:
\begin{equation}
  \partial_t(\rho\epsilon)+\partial_i(\rho\epsilon v_i)=
  \frac{\epsilon}{k}\left(c_{\epsilon 1}G_1+c_{\epsilon 2}G_2-c_{\epsilon 3}\rho\epsilon\right)+
  \partial_i(\rho c_\epsilon D\partial_i\epsilon),
\end{equation}
The proposed procedure of ``derivation'' should not be considered as
rigorous, the general aim of it was to show the correspondence between
terms of exact averaged hydrodynamic equations and model terms. For more
details see \cite{YanilkinStatsenkoKozlov, BHRZ}.

The full system of the model of turbulence (hereafter we drop the notation of averaging):
\begin{equation}
  \partial_t \rho + \partial_i(\rho v_i)=0,
  \label{eq:kesys_rho}
\end{equation}
\begin{equation}
  \partial_t(\rho v_i)+\partial_j\left(\rho v_i v_j+p\delta_{ij}\right)=-\partial_jR_{ij},
\end{equation}
\begin{equation}
  \partial_tE+\partial_i(v_i(E+p))=-G_2+\rho\epsilon+\partial_i(pa_i-Q_i^T),
\end{equation}
\begin{equation}
  R_{ij}=-\rho D \left(\partial_i v_j+\partial_j v_i-\frac{2}{3}\delta_{ij}\partial_kv_k\right)
  +\frac{2}{3}\rho k\delta_{ij},
\end{equation}
\begin{equation}
  \partial_t(\rho k)+\partial_i(\rho k v_i)=
  G_1+G_2-\rho\epsilon+\partial_i(\rho c_k D\partial_i k),
\end{equation}
\begin{equation}
  \partial_t(\rho\epsilon)+\partial_i(\rho\epsilon v_i)=
  \frac{\epsilon}{k}\left(c_{\epsilon 1}G_1+c_{\epsilon 2}G_2-c_{\epsilon 3}\rho\epsilon\right)+
  \partial_i(\rho c_\epsilon D\partial_i\epsilon),
\end{equation}
\begin{equation}
  E=\rho e + \frac{\rho v^2}{2},\quad
  D=c_D\frac{k^2}{\epsilon},\quad a_i=-c_\alpha D\frac{\partial_i\rho}{\rho},
\end{equation}
\begin{equation}
  G_1=-R_{ij}\partial_i v_j,\quad G_2=a_i\partial_i p,\quad
  Q_i^T=-c_e\rho D \partial_i e.
  \label{eq:kesys_G1G2Qi}
\end{equation}
One of the main weakness of the model are the unknown constants. The
procedure of model derivation does not fix the
constants. There are several ways to obtain their values: comparison with
experiments, the direct numerical simulation \cite{GuzhovaEtAl_VANT_2005}, some theoretical
approaches like renorm--group \cite{YakhotOrszag_JSciComp_1986}.
In this work the following set of constants is used \cite{GuzhovaEtAl_VANT_2005}: $c_\alpha=1.7$, $c_D=0.12$, $c_e=3$,
$c_{\epsilon 1}=1.15$, $c_{\epsilon 2}=1$, $c_{\epsilon 3}=1.7$,
$c_k=c_{\epsilon}=4/3$ (the model with the set of constants was tested
upon several laboratory experiments: Rayleigh--Taylor mixing
experiments \cite{DimonteSchneider_PhysFluids_2000},
Kelvin--Helmholtz mixing experiments
\cite{BrowandLatigo_PhysFluids_1979}, and direct numerical
simulations of these processes \cite{GuzhovaEtAl_VANT_2005}; the full
list of experiments is presented in \cite{GuzhovaEtAl_VANT_2005}).

To finish the model for burning we should add the influence of the turbulence on
flame (flame influences turbulence by generating gradients of
quantities). It is done in our work by changing the flame
velocity. In the regime when (\ref{eq:Ka_def}) is satisfied (the
flamelet regime), the effect of turbulence exhibits itself only in curvature
of flame surface. Such regime was considered in the paper \cite{Yakhot_CombSciTech_1988}
(see also \cite{Kerstein_CombSciTech_1988}) with a renorm--group
analysis. The result could be written as:
\begin{equation}
  \frac{v_{\rm turb}}{v_{\rm lam}}=\exp\left(\frac{2k}{v_{\rm turb}^2}\right),
  \label{eq:yakhot_vflame}
\end{equation}
where $v_{\rm lam}$ -- is a laminar speed (from
\cite{TimmesWoosley_ApJ_1992}), $v_{\rm turb}$ -- a turbulent flame
speed. The proposed expression is pure theoretical, though it was tested on
some experimental data (see \cite{Yakhot_CombSciTech_1988}), it may
need additional consideration.

\section{The problem setup}
\label{sec:problem}

We consider a white dwarf close to the Chandrasekhar
limit $M_{\rm Ch}$. We are interested in the process of a flame
propagation from the centre to outer regions of the WD:
according to evolutionary star models the flame is born near the
centre and, to meet observations, it should over time transform to the
detonation wave \cite{HillebrandtNiemeyer_astro_ph_0006305}.
The slow burning (flame) leads to expansion of matter, therefore in the region near the flame
the conditions of RTL instability growth are satisfied. On the
characteristic global timescale of flame propagation (here we mean $\sim
R_{\rm WD}/u_n$, not (\ref{eq:tau_flame})), RTL manages to evolve to
the nonlinear stage and develop turbulence. We are interested in its
intensity and its impact on the flame.

To answer these questions we use numerical simulations. Equations
(\ref{eq:kesys_rho})--(\ref{eq:kesys_G1G2Qi}) are implemented in our
numerical hydrocode FRONT3D \cite{front3d} in three-dimensional
case. As was mentioned earlier,
the model of turbulence can correctly reproduce 3D properties of
turbulence even in 1D simulations. Therefore to see the magnitude of
the turbulence effect we work in 1D spherical coordinates
in the scope of this paper. Not to encounter the problem of
building equilibrium WD configuration in Eulerian code, we implemented a lagrangian 1D
numerical scheme as a module in FRONT3D code for one-dimensional
simulations. We use an implicit scheme in mass coordinates
proposed in \cite{SamarskiiPopov} (the scheme contains the artificial
viscosity for shock wave problems, but our flows are significantly
subsonic, so it have no effect for our problem; furthermore, turbulence
leads to the appearance of the turbulent viscosity in our model, this
turbulent viscosity in much greater than artificial). This scheme uses staggered mesh: coordinate and
velocity are set at boundaries of cells $r_i$, $v_i$
and other quantities in the centres of cells $p_{i+1/2}$,
$\rho_{i+1/2}$ etc. Details of the implementation
can be found in the documentation that goes with the code. The
turbulent terms in hydrodynamic equations are included in this
scheme as external sources.

To treat correctly the properties of medium in the WD we use ``Helmholtz''
tabular equation of state \cite{helmholtz_eos}. The initial
distributions of all parameters are set to hydrodynamic equilibrium
with a numerical precision by the following procedure. The
distribution of mass $m_i$ and central density $\rho_c$ are set as a first step. Then
the recursive procedure determines profiles in the star:
\begin{equation}
  4\pi r_i^2\frac{p_{i+1/2}-p_{i-1/2}}{\Delta m_i}=-\frac{Gm_i}{r_i^2},
\end{equation}
\begin{equation}
  \rho_{i+1/2}={\rm EOS}(p_{i+1/2}, T_{i+1/2}),
\end{equation}
\begin{equation}
  r_{i+1}^3=r_i^3+\frac{3}{4\pi}\frac{\Delta m_{i+1/2}}{\rho_{i+1/2}}.
\end{equation}
This procedure requires the known $T(\rho,r)$ dependence.
The temperature after burning raises significantly, up to $\sim 10^{10}$~K, we could use any
small temperature as initial. Because thermoconductivity in WD is
strong, the star is usually isothermal in the centre. After all this, we set
constant $T_{\rm initial}$ everywhere. The choice of $T_{\rm initial}$
is presented below.
After integration we obtain the state of the WD close to the Emden
solution, but in equilibrium with the ``real'' EOS.

The Eq. (\ref{eq:G_eq}) describes evolution of flame surface for
general situation. For our case there exist much simpler procedure to
consider burning. The flame is defined by its position in mass
coordinates: $m_{\rm flame}$. The evolution of its coordinate
satisfies the equation:
\begin{equation}
  \frac{dm_{\rm flame}}{dt}=4\pi r_{\rm fl}^2\rho v_{\rm fl},
\end{equation}
where $r_{\rm fl}$ is the current position of flame, $\rho$, $v_{\rm fl}$ -- the density and
the flame velocity (as a function of medium state) at this point.
On every step energy released by burning is $\Delta Q=q\Delta m$,
$\Delta m$ is the mass of matter burned on the timestep (this energy
is spread on cells the flame moves throw).
The flame is initially set as a point $m_{\rm ign}$, from which two
fronts (forward and backward) start to propagate. The flame velocity
is calculated solving the implicit equation (\ref{eq:yakhot_vflame}) on each
time-step in our code for each flame front.

\section{Results}
\label{sec:results}

Here we present the results of simulations.
The parameters of the initial white dwarf are set as: the central
density $\rho_c=2\times 10^9$ g/cm$^3$, the initial composition as pure
${}^{12}$C or the mixture 0.5${}^{12}$C+0.5${}^{16}$O. The initial
temperature is calculated with the approximation for the ignition curve
from \cite{PotekhinChabrier_AstronAstrophys_2012}: for ${}^{12}$C it
is $T_{\rm initial}=2.7\times 10^{8}$~K, for
0.5${}^{12}$C+0.5${}^{16}$O it is $T_{\rm initial}=3.8\times 10^{8}$~K
(both temperatures are much smaller than the temperature after burning
$\sim 5\times 10^9$~K, what corresponds the statement from previous section).
We present results for several variants of the caloricity: $q_1=5.6\times
10^{17}$ erg/g (corresponds to transition C$\rightarrow$Mg), $q_2=9.2\times 10^{17}$ erg/g
(C$\rightarrow$Ni), and intermediate $q_3=7\times 10^{17}$ erg/g
(corresponds to transition to the NSE for carbon burning
\cite{NiemeyerHillebrandt_ApJ_1995}). With such variants of we
overlap a wide range of $q$ and check the dependence of results on it.
The conditions of convection preceding ignition are presented in the
paper \cite{NonakaEtAl_ApJ_2012}: our flame is ignited at the point
with $r_{\rm ign}=50$ km, initial turbulent velocity
$v_0''=\sqrt{2k_0}=16$ km/s, and the turbulent length $L_{\rm
  turb}=k_0^{3/2}/\epsilon_0=200$ km.

\begin{figure}[tbh]
  \begin{center}
    \includegraphics*[angle=270, width=0.6\linewidth]{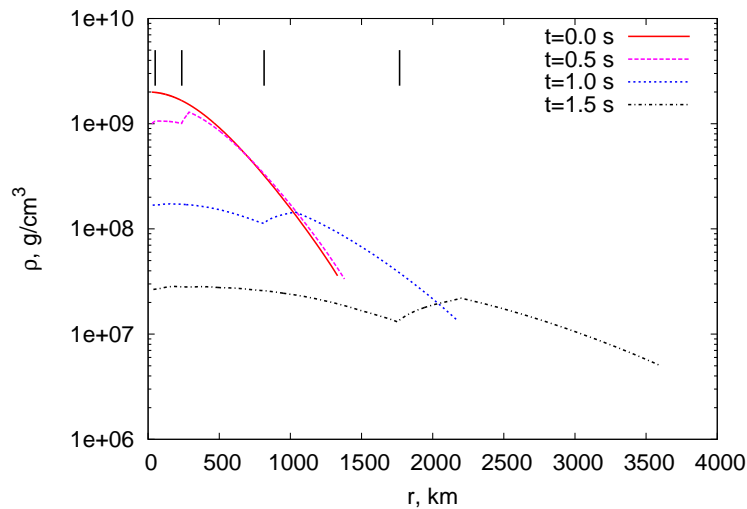}
    \caption{\label{fig:rho_ts}Density profiles for different moments,
    the variant with $q=9.2\times 10^{17}$ erg/g, ${}^{12}$C
    initial. Vertical marks show positions of the flame for these
    moments of time}
  \end{center}
\end{figure}

\begin{figure}[tbh]
  \begin{center}
    \includegraphics*[angle=270, width=0.6\linewidth]{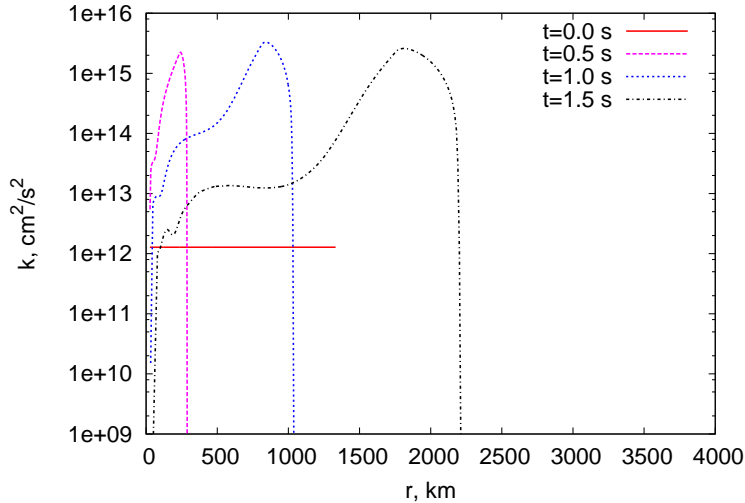}
    \caption{\label{fig:k_ts}Turbulent energy profiles for different moments,
    the variant with $q=9.2\times 10^{17}$ erg/g, ${}^{12}$C initial}
  \end{center}
\end{figure}

\begin{figure}
  \centering
  \subfloat{\label{fig:v_flames_v}  \includegraphics[angle=270,width=0.48\textwidth]{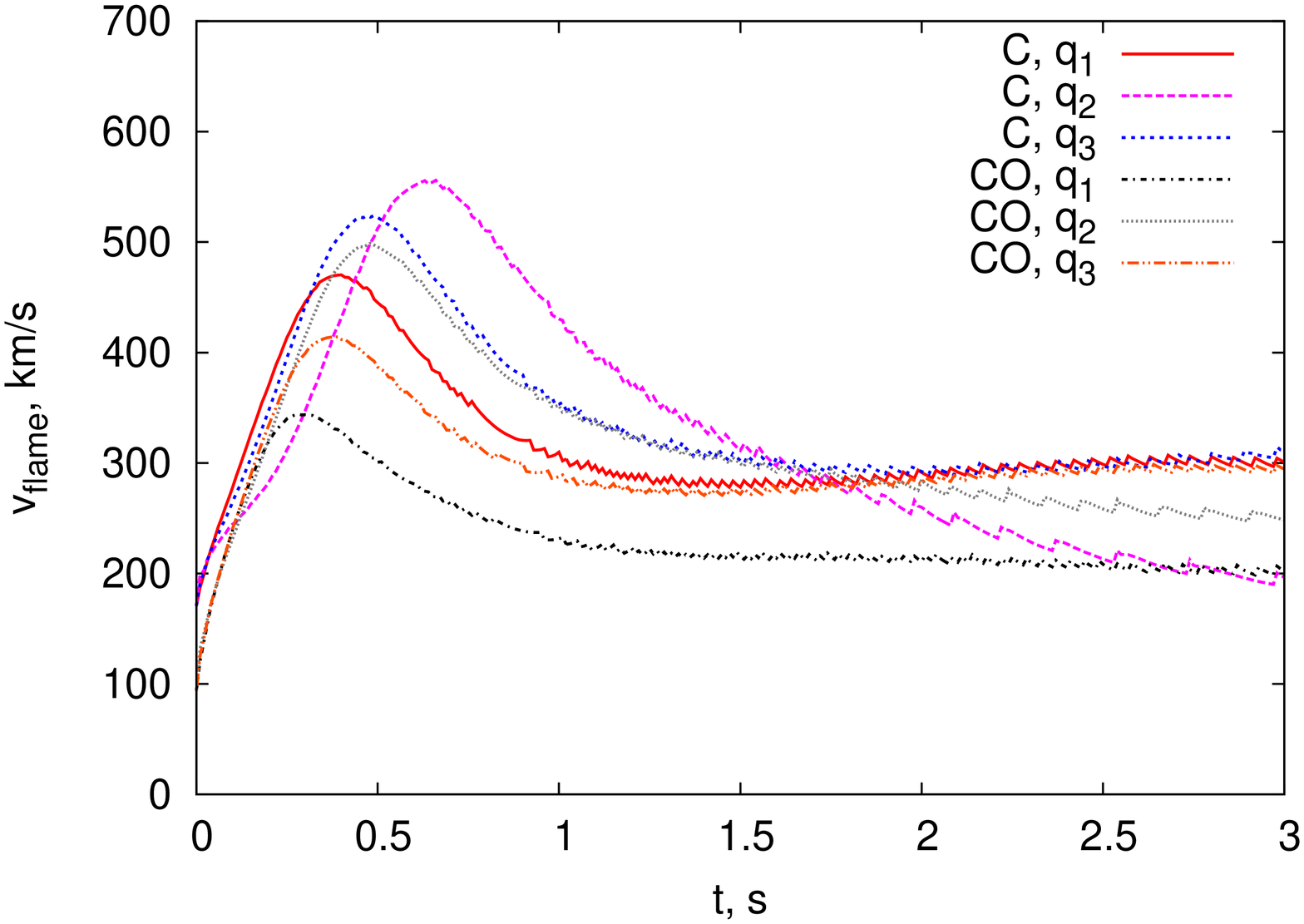}}
  \subfloat{\label{fig:v_flames_max}\includegraphics[angle=270,width=0.48\textwidth]{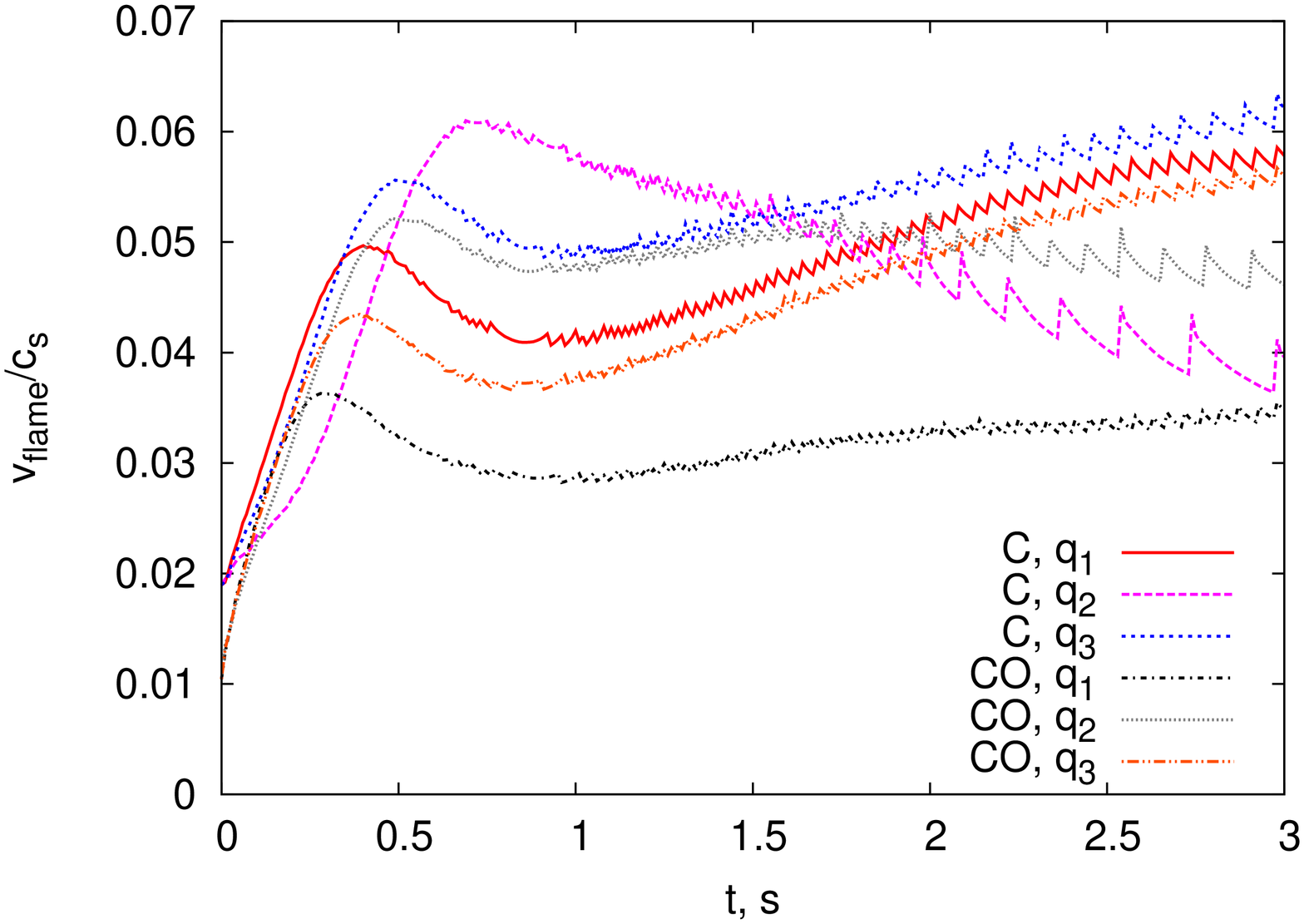}}
  \caption{Flame velocities (on the left) and the Mach numbers (on the
    right) for all variants}
  \label{fig:v_flames}
\end{figure}

The example of density evolution vs time is shown in Fig.~\ref{fig:rho_ts}.
The density drop, generated by flame spreading outwards, is smoothed
by the turbulence appeared.
The latter is maintained by the RTL instability, which growth is
proportional to the gravitational acceleration at the location of the
flame:
\begin{equation}
  g_{\rm fl}=\frac{G m_{\rm flame}}{r_{\rm flame}^2}.
  \label{eq:a_fl}
\end{equation}
Flame position $m_{\rm flame}$ grows with time together with $r_{\rm flame}$, the
cumulative effect of the instability yield is unknown a priori. The
profiles of specific turbulent
energy for the same simulation as in Fig.~\ref{fig:rho_ts} are shown
in Fig.~\ref{fig:k_ts}.
It could be seen that the maximum energy $\sim 2\times 10^{15}$ erg/g
(for the considered variant of initial conditions) is maintained with
time near the current position of the flame. This intensive turbulence
that exceeds background initial values is generated very early. That
is why the effect of changing initial turbulent
quantities $k_0$, $L_0$ have very small effect on the results, this
fact will be explicitly tested further, and shows the correctness of
the model work. The resulting turbulent
velocities of flame fronts for all variants are shown in Fig.~\ref{fig:v_flames}.
\begin{figure}[tbh]
  \begin{center}
    \includegraphics*[angle=270, width=0.6\linewidth]{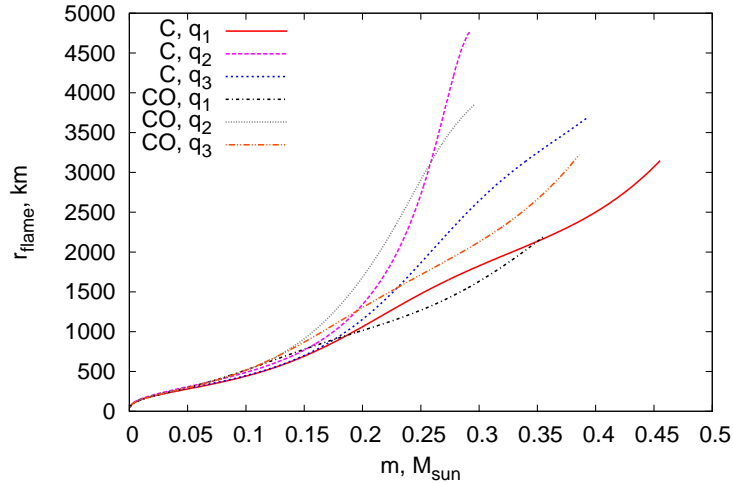}
    \caption{\label{fig:x_flames}Positions of the flame front versus
      massive coordinates for all variants}
  \end{center}
\end{figure}
\begin{figure}
  \begin{center}
    \includegraphics*[angle=270, width=0.6\linewidth]{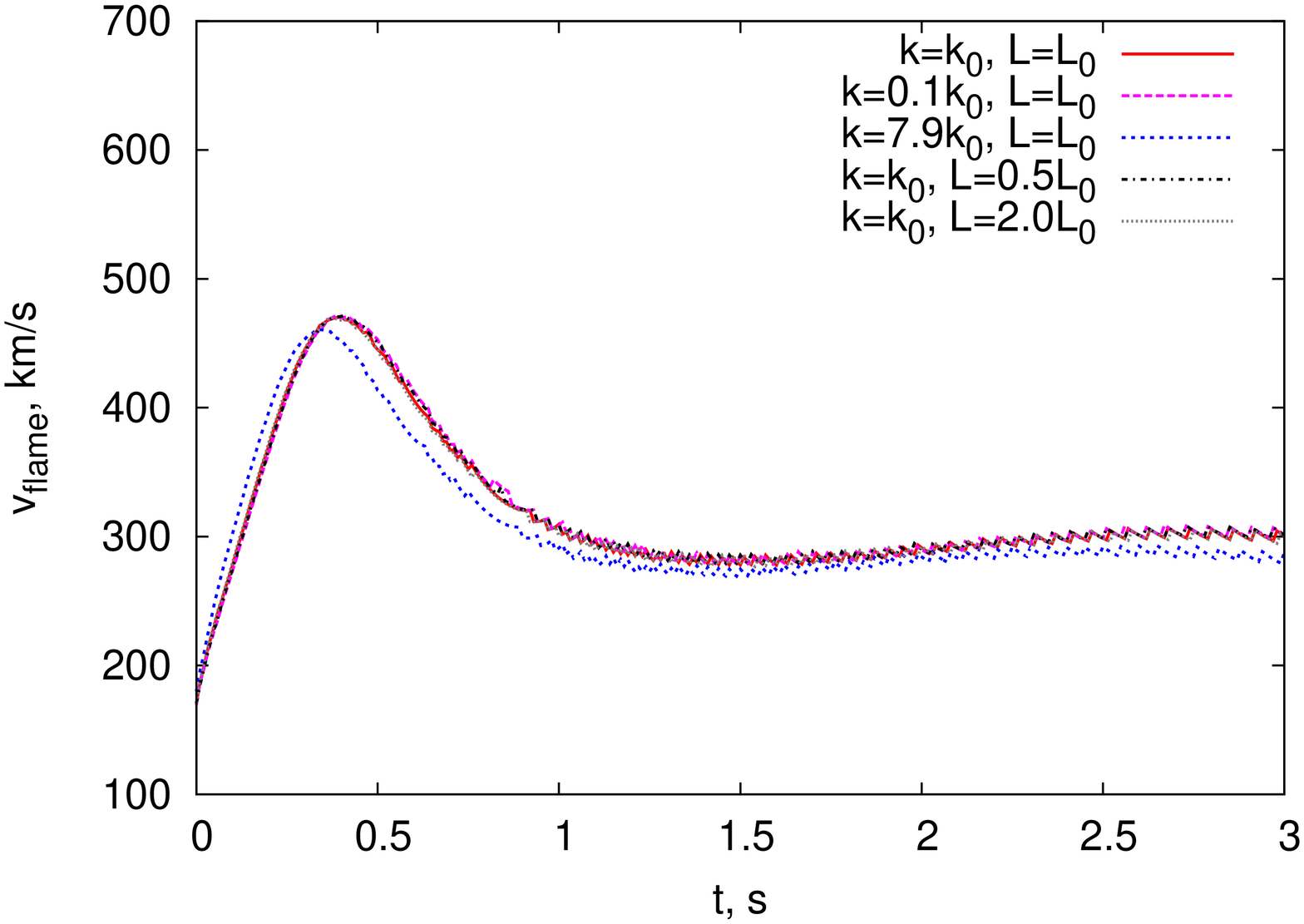}
    \caption{\label{fig:v_flames_c_0}Flame velocities for the variant
      with ${}^{12}$C and $q=q_1$ with various initial turbulent
      parameters $k_0$ and $L_0$}
  \end{center}
\end{figure}
The increase of velocity that occurs at early times is connected with
the raise of turbulent energy (without the turbulent impact the velocity
will monotonically decrease). Flame velocity satisfies
$v_{\rm flame}\approx \sqrt{2k}$ almost all the time, that is why more
precise relation instead of (\ref{eq:yakhot_vflame}) is not required
in these simulations, together with more exact $u_n(\rho, X_i)$
dependence. It could be seen that turbulence maintains flame
velocity on the level of (200-300) km/s, that is $\sim$ (3-7)\% of the
sound speed, what is much smaller than the latter and shows that this
turbulent regime does not lead to DDT. The Fig. \ref{fig:x_flames}
show the position of flame front versus massive coordinate for all
variants. It could be seen that larger caloricity leads to
faster expansion. Relatively large velocity of turbulent flame
maintains active burning till the end of simulation, when $\rho$ drops to
several $10^6$~g/cm$^3$.

Additional runs were made with increased (or decreased) values of
$k_0$, $L_{\rm turb_0}$ (what correspond to the variation of initial
turbulence), showing the independence of final turbulence
on the variation of initial and background values, see Fig.~\ref{fig:v_flames_c_0}.

\section{Conclusions}
\label{sec:conclusions}

The paper considers the problem of a white dwarf burning with account
for turbulence. The core of the approach is a $k$-$\epsilon$
model of turbulence, which succeeds in turbulence simulations in any
dimension: 1-, 2-, 3D, reproducing correctly 3D turbulence
properties. Using this approach we presented 1D simulations of a white dwarf for
${\rm Ka}\ll 1$ regime of turbulent burning (the flamelet regime). As a result the stationary
turbulent intensity arises after some time in simulations. This
turbulence maintains flame velocity at $v_{\rm flame}\approx (250\pm
50)$~km/s. The obtained turbulent velocities agree with the results
of more sophisticated simulations like \cite{Ropke_ApJ_2007}. The
flame acceleration (the final Mach number) $\sim 0.05 c_s$ is small and we can conclude, similar to
\cite{WoosleyKersteinEtAl_ApJ_2009}, that deflagration to detonation
transition occurs in different regime of turbulent burning (not
in the flamelet), or on the border of two regimes (there could be identified
three regimes of turbulent burning -- the flamelet, the stirred flame,
the well-stirred flame, for additional details see
\cite{WoosleyKersteinEtAl_ApJ_2009}) of turbulent flame
propagation.
It is important to note that this conclusion refers to the whole
flame, according to \cite{Ropke_ApJ_2007, SchmidtEtAl_ApJ_2010} the
detonation can be triggered by rare high-velocity turbulent
fluctuations. Such fluctuations are not reproduced by the proposed
model correctly, it needs some modifications.

This work is planned to be continued with simulations in higher dimensions --
2D, 3D, studying of other regimes of turbulent flame, implementation
of more sophisticated model of nuclear reactions, and comparison with observations.
Also the considered model was tested upon terrestrial experiments on
turbulent mixing (see the Section~\ref{sec:model}), the question of its
application for SNIa turbulence is not fully clear and should be considered further. Nevertheless such
type of a model have benefits in reproducing turbulent properties in
lower dimensional simulations and could be used for the low-cost studies.

The author is grateful to S. Blinnikov, F. R\"opke for discussions,
an anonymous referee for very valuable comments. The work is
also supported by RFBR grants 11-02-00441-a and 13-02-92119,
Sci. Schools 5440.2012.2, 3205.2012.2 and 3172.2012.2,
 by the contract No.~11.G34.31.0047 of the Ministry of Education and
 Science of the Russian Federation, and SCOPES project No.~IZ73Z0-128180/1.

\bibliographystyle{apsrev}
\bibliography{turb_wd1d_keps}

\end{document}